\documentclass[journal,comsoc]{IEEEtran}
\usepackage{cite}
\usepackage{amsmath,amssymb,amsfonts}
\usepackage{graphicx}
\usepackage{textcomp}
\usepackage{xcolor}
\usepackage{caption}
\usepackage{subcaption}
\usepackage{kotex}
\usepackage{amsthm}
\usepackage{algpseudocode,algorithm,algorithmicx}

\theoremstyle{definition}

\newtheorem{lemma}{Lemma}

\begin{document}
\title{Impacts of Device Caching of Content Fractions on Expected Content Quality}

\author{Dongjae Kim,~\IEEEmembership{Member,~IEEE,}
	and~Minseok Choi,~\IEEEmembership{Member,~IEEE}% <-this % stops a space
	\thanks{Dongjae Kim is with the Artificial Intelligence Convergence Research Center for Regional Innovation, Korea Maritime \& Ocean University, Busan 49112, South Korea (e-mail: kdj6306@gmail.com).}
	\thanks{Minseok Choi is with the Department of Telecommunication Engineering, Jeju National University, Jeju 63243, South Korea (e-mail: ejaqmf@jejunu.ac.kr)}}

\markboth{Submission to IEEE Wireless Communication Letters}
{Shell \MakeLowercase{\textit{et al.}}: Bare Demo of IEEEtran.cls for IEEE Communications Society Journals}

\maketitle
\begin{abstract}
	This paper explores caching of fractions of a video content, not caching of an entire content, to increase the expected video quality. 
	We first show that the highest-quality content is better to be cached and propose the caching policy of video chunks having different qualities.
	Our caching policy utilizes the characteristics of video contents that video files can be encoded into multiple versions with different qualities, each file consists of many chunks, and chunks can have different qualities. 
	Extensive performance evaluations are conducted to show that caching of content fractions, rather than an entire content, can improve the expected video quality especially when the channel conditions is sufficiently good to cooperate with nearby BS or helpers. 
\end{abstract}


\IEEEpeerreviewmaketitle

\section{Introduction}

\IEEEPARstart{A}{s} the global data traffic is increasingly exploding, the wireless caching network \cite{TITShanmugam} has been considered as a key technology to deal with the repeated and overlapped requests of popular video contents which account for the largest proportion of the global data traffic \cite{TMM2013Cheng}. 
Caching popular contents on the finite storage of the caching helper was proposed to reduce the delivery latency \cite{TITShanmugam}, and a device-to-device (D2D)-assisted caching network has been also studied \cite{JSAC2016Ji}. 
Further, caching and delivery methods optimized for video contents or streaming services have been developed with consideration of characteristics of a video content.

A video content or a stream has several features that affect the caching and delivery schemes.
First, a video content can be encoded to multiple versions having different qualities, e.g., bitrates, and the content size depends on its quality level; therefore, it is very important to decide which content and what quality to cache and deliver depending on the network status \cite{JSAC2018Choi}.
Second, a video file consists of many chunks that are in charge of the fixed playtime. 
%The dynamic delivery of video chunks depending on the channel and cache states have been proposed in \cite{TWC2019Choi}. 
Third, dynamic video streaming allows each chunk to have
a different quality. %and some researchers addressed video delivery and scheduling methods which support seamless streaming by dynamically selecting the quality level \cite{JCN2021Choi}. 
This paper takes the above characteristics of a video content in to account for designing the caching method, even though there are some more researches considering video-specific features, e.g., joint optimization of content recommendation and caching \cite{TMC2019Chat}, and temporal correlations among consecutive content requests \cite{TWC2021Choi}.

Caching of contents having different qualities is developed in \cite{JSAC2018Choi} and \cite{TMC2020Qu}, and the adaptive bitrate video caching and processing is presented in \cite{TMC2019Tran}; however, they do not consider the feature that a video content can be divided into multiple chunks. 
The authors of \cite{TWC2019Choi} and \cite{JCN2021Choi} proposed dynamic delivery of video chunks having different qualities assuming that popular contents are entirely cached. 
Caching of partial files is presented in \cite{NCC2019Narayana} when popularity of video segments is different; however, different quality levels for segments are not considered.
The goal of this paper is to develop the policy of caching content fractions (e.g., video chunks) and to analyze the impact of caching content fractions on the expected video quality. 
This policy determines 1) how much of the content fractions and 2) which quality level to be cached in cache-enabled devices to maximize the expected quality with consideration of the time-varying channel condition and random device distributions. 

Motivations of caching of content fractions are as follows:
\begin{itemize}
	\item Fragmentation of a content is common in the practical scenario, e.g., chunks or segments in video streaming. 
	
	\item Most of the existing studies on wireless caching assume an identical content size because a large one can be divided into multiple pieces \cite{TITShanmugam,JSAC2016Ji}; however, they overlook the fact that a user should collect all of the pieces to enjoy the content. 
	This paper considers the recovery of the whole pieces of the desired content at the user side. 
	
	\item When the content size is very large, caching becomes too biased, i.e., few of the most popular and/or low-quality contents would occupy the limited storage size. 
	If the channel condition is sufficiently good, caching of fractions of high-quality popular contents would be better for achieving high quality and the fairness among contents by cooperation of other caching entities that can deliver uncached fractions. 
\end{itemize}

\vspace{-2mm}
\section{System Model}
\label{sec:system_and_problem}
%\subsection{System Model}
%\label{subsec:system}

This paper considers a wireless caching network with a base station (BS) and cache-enabled devices distributed within the radius $R$ of the BS. 
Devices randomly request contents from a library $\mathcal{F}$ consisting of $F$ contents.
%Suppose that all the contents have $L$ quality levels, and the quality measure of the content with the $l$-th quality level is denoted as $q_l$.
Suppose that contents can be encoded into different quality levels within the interval of $q \in [q_{\text{min}}, q_{\text{max}}]$\footnote{In practice, there are finite quality levels that contents can be encoded to; however, a continuous quality interval is considered in this letter for mathematical convenience, and we can address that the proposed caching method shows the upper bound on the performance in the practical scenario.}. 
The size of a content is proportional to $q$, denoted as $S(q) = A \cdot q$ where $A$ is a coefficient. 
Also, all contents have to be delivered to a content-requesting device until time $T$. 
We consider contents, e.g., video files, which can be divided into some fractions and each fraction can have a different quality level. %\footnote{For example, a video stream in the DASH system \cite{DASH} consists of several chunks that are in charge of the fixed playtime. In adaptive bitrate streaming, chunks in a stream can have different bitrates and users' quality of experiences (QoE) is affected by the average bitrate and its variations \cite{JSAC2018Choi,TWC2019Choi}}. 
Different from most of the existing studies \cite{TITShanmugam,JSAC2016Ji, JSAC2018Choi,TWC2019Choi,TWC2021Choi}, we allow caching of content fractions in cache-enabled devices. 
We also assume that each device determines only one quality level $q_i$ for caching a fraction $\alpha_i$ of the content $i$.
Then, the policy of caching content fractions can be denoted as $\{\boldsymbol{\alpha, \mathbf{q}} \}$, where $\boldsymbol{\alpha} = [\alpha_1 ,\cdots,\alpha_F]$, $\mathbf{q} = [q_1 ,\cdots,q_F]$, where $\alpha_i$ is a fraction of the content $i$ to be cached in devices and $q_i$ is the quality level of the cached content fraction. 
Let the device have a finite storage size of $M$, then the following inequality is satisfied if the policy $\{\boldsymbol{\alpha}, \mathbf{q} \}$ is adopted:
\begin{equation}
A \sum_{i=1}^F \alpha_i q_i \leq M.
\label{eq:cache_const}
\end{equation} 
If a device requests the content $i$, $\alpha_i$ of the content can be directly obtained from its storage and the remaining $(1-\alpha_i)$ should be delivered from other cache-enabled entities.
If the quality of the delivered fraction is $q_0$, then we let the expected content quality that the user enjoys be $\alpha_i q_i + (1-\alpha_i) q_0$. 

In order to design the policy $\{\boldsymbol{\alpha, \mathbf{q}} \}$ and to analyze the impact of caching content fractions, we consider a simple scenario in which only a BS can deliver the remaining fraction of the desired content. 
Here, the BS has access to the server in the core network having the whole file library; therefore, a cache hit occurs whatever the user requests, but the problem is to determine how much of the contents to cache to devices and how much quality of the uncached content fraction can be provided from the BS. 
We suppose that the BS is deployed with the edge computing server so that it can transcode the content to complete the delivery until the time $T$ depending on the channel capacity.

%Suppose that devices are uniformly distributed within the radius $R$ of a BS and an identical frequency band $\mathcal{B}$ is allocated to devices. 
The Rayleigh fading channel between the BS and a device is denoted by $\tilde{h}(t) = \sqrt{D(t)}h(t)$ at time $t$, where $D(t) = 1/r^{\beta}$ denotes the inverse of the path loss, $r$ and $\beta$ are the distance between the BS and the device and the path loss exponent, respectively, and $h(t)\sim\mathcal{CN}(0,1)$ represents a fast fading component. 
Here, we assume a block fading model where $\tilde{h}(t)$ is static during the time $T$, and that the shadowing effects are ignored for mathematical convenience.
We adopt the distance-based interference control \cite{TWC2019Choi}, and assume that the BS can communicate with the device within the radius $R$ with the fixed interference-to-noise ratio (INR) $\Upsilon$; therefore, the channel capacity is represented as $C = \mathcal{B} \log_2 (1+ \Psi |\tilde{h}|^2/(\Upsilon + 1))$, where $\Psi$ is the transmit signal-to-noise ratio (SNR).

%\subsection{Problem Formulation}
When the device requests the content $i$, the BS delivers the uncached fraction $(1-\alpha_i)$ of the content $i$ to the device; however, its quality depends on the channel capacity. 
The BS is willing to transmit the content fraction with the highest quality as long as the channel capacity allows, and the transcoded quality becomes $\frac{CT}{A(1-\alpha_i)}$.
Then, the expected quality of the device is defined as 
\begin{equation}
\sum_{i=1}^F f_i \left[\alpha_i q_i+(1-\alpha_i)\mathbb{E}\left[\min \left\{\frac{CT}{A(1-\alpha_i)},q_{\max} \right\} \right]\right],
\label{eq:expected_qual}
\end{equation}
where $f_i$ is the popularity of the content $i$.
In \eqref{eq:expected_qual}, the first term represents the quality of cached fractions and the second is the expected quality of uncached fractions that the BS transmits.

\section{Optimal Caching Policy for a Device with Fixed Location}
\label{sec:fixed}

In this section, the optimal caching policy is proposed for a device with a fixed location.
The optimization problem maximizing the expected quality is mathematically formulated as
\begin{align}\label{eq:opt problem for fixed1}
&~~~~\max_{\boldsymbol{\alpha},\mathbf{q}} \sum_{i=1}^{F} f_i g(\alpha_i,q_i)\\
\text{s.t.}~ \eqref{eq:cache_const}&,~ %\sum_{i=1}^F \alpha_i q_i \leq M&,~
0 \leq \alpha_i \leq 1,~
q_{\min} \leq q_i \leq q_{\max}, ~ \forall{i}
\end{align}
where $g(\alpha_i,q_i)$ is defined as
\begin{equation}
g(\alpha_i,q_i)=\alpha_i q_i +  \mathbb{E}_h \left[\min \left\{\frac{CT}{A},q_{\max} (1-\alpha_i) \right\} \right].
\end{equation}
By substituting $x_i=\alpha_i q_i$, $g$ can be expressed as a function of $\alpha_i$ and $x_i$ as below
\begin{equation}\label{eq:g function wrt x}
g\left(\alpha_i,\frac{x_i}{\alpha_i}\right)=x_i +  \mathbb{E}_h \left[\min \left\{\frac{CT}{A},q_{\max} (1-\alpha_i) \right\} \right].
\end{equation}
Thus, the problem in \eqref{eq:opt problem for fixed1} can be converted into
\begin{align}\label{eq:opt problem for fixed2}
&\max_{\boldsymbol{\alpha},\mathbf{x}} \sum_{i=1}^{F} f_i g\left(\alpha_i,\frac{x_i}{\alpha_i}\right)\\
\text{s.t.}~ \sum_{i=1}^F x_i \leq \frac{M}{A}&,~
0 \leq \alpha_i \leq 1,~
q_{\min} \leq \frac{x_i}{\alpha_i} \leq q_{\max}, ~ \forall{i}. \label{eq:constraints for fixed2}
\end{align}
First, we propose the optimal $\alpha_i$ for given $x_i$ in the following lemma.
We prove that $g(\alpha_i,\frac{x_i}{\alpha_i})$ is a decreasing function with respect to $\alpha_i$. 
Then, considering the constraints \eqref{eq:constraints for fixed2}, the optimal $\alpha_i$ for given $x_i$ can be obtained.

\begin{lemma}
	\label{lemma:lemma1}
	When the distance $r$ between BS and device is fixed, the optimal $\alpha_i$ is $\alpha_i=x_i/q_{\max}$ for given $x_i$.
	\begin{proof}
		Let $u$ be the channel power $u=|h|^2$, then $u$ has an exponential distribution such that $f_U(u)=e^{-u}$ for $u\geq 0$, and
		%\begin{equation}
		%    f_U(u)=e^{-u},~~\text{for}~u\geq0.
		%\end{equation}
		the channel capacity is a function of $u$ as $C(u)=\mathcal{B}\log_2\big(1+\frac{\Psi u}{d^\beta(\Upsilon+1)}\big)$.
		The expectation in \eqref{eq:g function wrt x} is derived as
		\begin{align}
		&\mathbb{E}_h \left[\min \left\{\frac{C(u)T}{A},q_{\max} (1-\alpha_i) \right\} \right]\notag \\
		&~~~=\int_0^\infty\min \left\{\frac{C(u)T}{A},q_{\max} (1-\alpha_i) \right\}e^{-u} du \notag \\
		&~~~=\int_0^{k(\alpha_i)}\frac{C(u)T}{A}e^{-u} du
		+\int_{k(\alpha_i)}^\infty q_{\max}(1-\alpha_i) e^{-u} du,
		\end{align}
		where $k(\alpha_i)=r^{\beta}\frac{\Upsilon+1}{\Psi} \Big(2^{\frac{Aq_{\max}(1-\alpha_i)}{T\mathcal{B}}}-1\Big)$, which is obtained from $C(k(\alpha_i))T/A=q_{\max}(1-\alpha_i)$. The partial derivative of $g\big(\alpha_i,\frac{x_i}{\alpha_i}\big)$ with respect to $\alpha_i$ is
		\begin{align}
		&\frac{\partial}{\partial \alpha_i}g\Big(\alpha_i,\frac{x_i}{\alpha_i}\Big)
		=\frac{\partial}{\partial \alpha_i}\mathbb{E}_h \left[\min \left\{\frac{C(u)T}{A},q_{\max} (1-\alpha_i) \right\} \right]\notag\\
		&=\frac{C(k(\alpha_i))T}{A}e^{-k(\alpha_i)}\frac{\partial k(\alpha_i)}{\partial  \alpha_i}-q_{\max}(1-\alpha_i)e^{-k(\alpha_i)}\frac{\partial k(\alpha_i)}{\partial \alpha_i}\notag\\
		&~~~-q_{\max}e^{-k(\alpha_i)}=-q_{\max}e^{-k(\alpha_i)}<0.
		\end{align}
		Thus, $g\big(\alpha_i,\frac{x_i}{\alpha_i}\big)$ is a decreasing function for $\alpha_i$. From the constraints \eqref{eq:constraints for fixed2}, $\alpha_i \geq x_i/q_{\max}$. Therefore, the optimal $\alpha^*_i=x_i/q_{\max}$.
	\end{proof}
\end{lemma}
Since $q_i=x_i/\alpha_i$ from the definition of $x_i$, the optimal $q^*_i=q_{\max}$. 
It means that given $x_i=\alpha_i q_i$, increasing quality $q_i$ is better for achieving the high expected quality than caching more fractions $\alpha_i$.

Now, we concentrate on obtaining $x_i$. 
Using the Lemma \ref{lemma:lemma1}, $g\big(\alpha_i,\frac{x_i}{\alpha_i}\big)$ is rewritten as $g\big(\frac{x_i}{q_{\max}},q_{\max}\big)$ by substituting $\alpha_i=x_i/q_{\max}$.
The partial derivative of $g\big(\frac{x_i}{q_{\max}},q_{\max}\big)$ with respect to $x_i$ can be derived as
\begin{equation}\label{eq:derivative g wrt x}
\frac{\partial}{\partial x_i}g\left(\frac{x_i}{q_{\max}},q_{\max}\right)=1-e^{-k\big( \frac{x_i}{q_{\max}}\big)}.
\end{equation}
Since $k\big(\frac{x_i}{q_{\max}}\big)>0$, the derivative \eqref{eq:derivative g wrt x} is positive so that $g\left(\frac{x_i}{q_{\max}},q_{\max}\right)$ is an increasing function of $x_i$.
Therefore, the expected quality increases as $x_i$ grows.

\begin{figure}[t]
	\centering
	\includegraphics[width=0.4\textwidth, trim=-0.5cm 0 0 0]{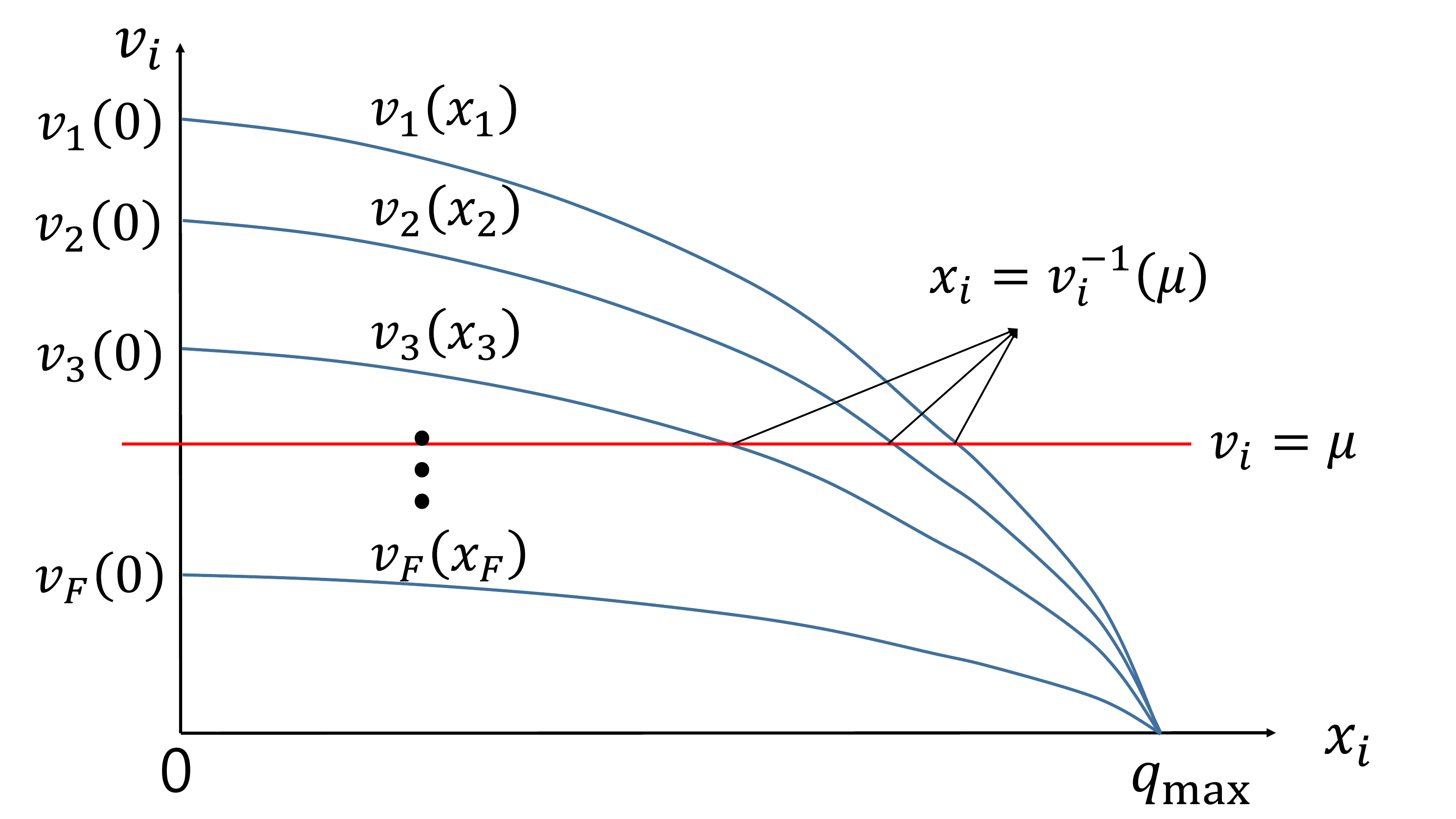}
	\caption{Increment rate of the expected quality versus $x_i$.}\label{fig:increase rate}
\end{figure}

From now on, how to find $\left\{x_i\right\}$ that maximizes the expected quality while satisfying constraint \eqref{eq:constraints for fixed2} is described.
For the content $i$, let $v_i(x_i)$ be the increment rate of the expected quality with respect to $x_i$. Then $v_i(x_i)$ is expressed as 
\begin{equation}
v_i(x_i)=f_i\frac{\partial}{\partial x_i}g\left(\frac{x_i}{q_{\max}},q_{\max}\right)
=f_i\left(1-e^{-k\big(\frac{x_i}{q_{\max}}\big)}\right).
\end{equation}
Since $k\big(\frac{x_i}{q_{\max}}\big)$ is a decreasing function of $x_i$, $v_i(x_i)$ is also a decreasing function of $x_i$.
Without loss of generality, it is assumed that the popularity $\left\{f_i\right\}$ are arranged in descending order.
Then, the increment rate $v_i(x_i)$ for all contents $i \in \{1,\cdots,F\}$ can be shown as Fig. \ref{fig:increase rate}. 
If $v_i(x_i)>v_j(x_j)$, increasing $x_i$ is better for growing the expected quality than increasing $x_j$.
In Fig. \ref{fig:increase rate}, the constraint of the cache size in \eqref{eq:cache_const} can be reflected by using a predetermined threshold $\mu$.
As long as $\mu$ allows, all contents try to increase their $x_i$ values as much as possible in the order in which $v_i(x_i)$ is large, and their maximum value becomes $x_i = v_i^{-1}(\mu)$.
%Given $\mu$, for all content indices $\left\{i\right\}$ such that $v_i(0)\geq \mu$, the optimal $x^*_i$ satisfies constraints \eqref{eq:constraints for fixed2} while $v_i(x_i)$ has the same value $\mu$.
As $\mu$ decreases, $x_i=v_i^{-1}(\mu)$ gradually increases; therefore, we can find the appropriate $\mu^*$ which satisfies $\sum_{i=1}^F x_i=M/A$, and $x^*_i = v_i^{-1}(\mu^*)$ for all $i\in\{1,2,\cdots, F \}$ become optimal.
Here, if the popularity $f_i$ is very low and $v_i(0)<\mu$, this content $i$ cannot be cached at all.
This process of finding $x^*_i$ is similar to the water-filling method, and the details are described in Algorithm \ref{alg:water-filling 1} by using the bisection method. 
Finally, the optimal $\alpha^*_i$ can be obtained as $\alpha^*_i=x^*_i/q_{\max}$ using Lemma \ref{lemma:lemma1}.

\begin{algorithm}[t]
	\caption{Algorithm to find $\left\{x^*_i\right\}$ in a fixed-distance scenario}\label{alg:water-filling 1}
	\begin{algorithmic}
		\State \textbf{Initialize} $a=0$ and $b=v_1(0)$
		\While{$b-a \geq \epsilon$}
		\State{$\mu=\frac{a+b}{2}$}
		\For{$i=1,\cdots,F$}
		\If{$\mu>v_i(0)$}
		\State $x_i=0$
		\ElsIf{$\mu \leq v_i(0)$}
		\State $x_i=v_i^{-1}(\mu)$
		\EndIf
		\EndFor
		\If{$\sum_{i=1}^F x_i > M/A$}
		\State $a=\mu$
		\ElsIf{$\sum_{i=1}^F x_i \leq M/A$}
		\State $b=\mu$
		\EndIf
		\EndWhile
	\end{algorithmic}
\end{algorithm}
\vspace{-2mm}

\section{Optimal Caching Policy for a Randomly Distributed Device}

In this section, the optimal caching policy is proposed when user distance $r$ is randomly distributed. 
Here, we assume that the user is uniformly distributed within the radius is $R$ of the BS having an access to the whole library $\mathcal{F}$.
The probability density function of $r$ is $f_R(r)=2r/R^2$ for $0\leq r \leq R$.
%\begin{equation}
%    f_R(r)=\frac{2r}{R^2}, ~ \text{for}~ 0\leq r \leq R.
%\end{equation}
Similarly, the expected quality of the content $i$ at the user side is defined as
\begin{equation}
g(\alpha_i,q_i)=\alpha_i q_i +  \mathbb{E}_{h,r} \left[\min \left\{\frac{CT}{A},q_{\max} (1-\alpha_i) \right\} \right]. 
\label{eq:g_random}
\end{equation}
Here, the expectation in \eqref{eq:g_random} is with respect to the distance $r$ as well as the channel gain $h$.
As explained in Section \ref{sec:fixed}, $g$ can be expressed as a function of $\alpha_i$ and $x_i$ by substituting $x_i=\alpha_i q_i$ as follows:
\begin{equation}\label{eq:g function wrt x2}
g\left(\alpha_i,\frac{x_i}{\alpha_i}\right)=x_i +  \mathbb{E}_{h,r} \left[\min \left\{\frac{CT}{A},q_{\max} (1-\alpha_i) \right\} \right].
\end{equation}
Then, we can prove that $q^*_i=q_{\text{max}}$ always and find the relationship between $\alpha_i$ and $x_i$ by using the following lemma. 

\begin{lemma}
	\label{lemma:lemma2}
	When a content-requesting device is uniformly distributed within the radius $R$ of the BS, the optimal $\alpha^*_i=x_i/q_{\max}$ for given $x_i$.
	\begin{proof}
		Due to the randomness of $r$, the channel capacity is a function of both $u=|h|^2$ and $r$ as $C(u,r)=\mathcal{B}\log_2\big(1+\frac{\Psi u}{r^\beta(\Upsilon+1)}\big)$. Then, the expectation in \eqref{eq:g function wrt x2} can be derived as follows:
		\begin{align}
		\mathbb{E}_{h,r} &\left[\min \left\{\frac{C(u,r)T}{A},q_{\max} (1-\alpha_i) \right\} \right]\notag \\
		&=\int_0^R\int_0^\infty\min \left\{\frac{C(u,r)T}{A},q_{\max} (1-\alpha_i) \right\}e^{-u}du f_R(r)dr \notag \\
		&=\int_0^R\Bigg[\int_0^{l(\alpha_i)r^\beta}\frac{C(u,r)T}{A}e^{-u} du\notag\\
		&~~~~~~~~~~~~~~~+\int_{l(\alpha_i)r^\beta}^\infty q_{\max}(1-\alpha_i) e^{-u} du\Bigg]f_R(r)dr,
		\end{align}
		where $l(\alpha_i)=\frac{\Upsilon+1}{\Psi} \Big(2^{\frac{Aq_{\max}(1-\alpha_i)}{T\mathcal{B}}}-1\Big)$ which is obtained from $C(l(\alpha_i)r^\beta,r)T/A=q_{\max}(1-\alpha_i)$.
		The partial derivative of $g\big(\alpha_i,\frac{x_i}{\alpha_i}\big)$ with respect to $\alpha_i$ is
		\begin{align}
		\frac{\partial}{\partial \alpha_i}g\left(\alpha_i,\frac{x_i}{\alpha_i}\right)
		&=\frac{\partial}{\partial \alpha_i}\mathbb{E}_{h,r} \left[\min \left\{\frac{C(u,r)T}{A},q_{\max} (1-\alpha_i) \right\} \right]\notag\\
		&=-\int_{0}^{R} e^{-l(\alpha_i)r^\beta} f_R(r) dr<0.
		\end{align}
		Thus, $g\big(\alpha_i,\frac{x_i}{\alpha_i}\big)$ is a decreasing function for $\alpha_i$. 
		From the constraints \eqref{eq:constraints for fixed2}, $\alpha_i \geq \frac{x_i}{q_{\max}}$. Therefore, the optimal $\alpha^*_i={x_i}/{q_{\max}}$ given $x_i$.
	\end{proof}
\end{lemma}

Same as in Section \ref{sec:fixed}, the optimal quality becomes $q^*_i=x_i/\alpha^*_i = q_{\text{max}}$ for all $i$, which does not depend on the distribution of $r$.
%This result is the same even if the distribution $f_R(r)$ is not uniform.
Accordingly, whatever distribution $f_R(R)$ is, increasing the quality $q_i$ rather than caching more fractions $\alpha_i$ is beneficial for achieving the high expected quality.

Now, we concentrate on obtaining the optimal $x^*_i$. 
Using the Lemma \ref{lemma:lemma2}, $g\big(\alpha_i,\frac{x_i}{\alpha_i}\big)$ is rewritten as $g\big(\frac{x_i}{q_{\max}},q_{\max}\big)$ by substituting $\alpha_i={x_i}/{q_{\max}}$.
The partial derivative of $g\big(\frac{x_i}{q_{\max}},q_{\max}\big)$ with respect to $x_i$ can be derived as
\begin{equation}\label{eq:derivative g wrt x rd}
\frac{\partial}{\partial x_i}g\left(\frac{x_i}{q_{\max}},q_{\max}\right)=1-\int_0^Re^{-l(\frac{x_i}{q_{\max}})r^\beta} f_R(r) dr.
\end{equation}
Since $l\big(\frac{x_i}{q_{\max}}\big)>0$, the derivative \eqref{eq:derivative g wrt x rd} is positive, that is, $g\big(\frac{x_i}{q_{\max}},q_{\max}\big)$ is an increasing function of $x_i$.
Therefore, the expected quality increases as $x_i$ grows.
Now, the integral in \eqref{eq:derivative g wrt x rd} can be computed as follows:
\begin{align}
\int_0^Re^{-l\big(\frac{x_i}{q_{\max}}\big)r^\beta} f_R(r) dr
= \frac{2}{\beta R^2 l\big(\frac{x_i}{q_{\max}}\big)^{\frac{2}{\beta}}} \gamma \left(\frac{2}{\beta}, l\big(\frac{x_i}{q_{\max}}\big)R^\beta \right)
\end{align}
where $\gamma(s,t)$ is the lower incomplete gamma function $\gamma(s,t)=\int_0^t e^{-z} z^{s-1} dz$.
Through this, the increment of the expected quality for $x_i$ can be expressed as 
\begin{align}
v_i(x_i)&=f_i \frac{\partial}{\partial x_i}g\left(\frac{x_i}{q_{\max}},q_{\max}\right) \notag\\
&=f_i \Bigg[1-\frac{2}{\beta R^2 l\big(\frac{x_i}{q_{\max}}\big)^{\frac{2}{\beta}}} \gamma \left(\frac{2}{\beta}, l\big(\frac{x_i}{q_{\max}}\big)R^\beta \right) \Bigg]
\end{align}
Since $l\big(\frac{x_i}{q_{\max}}\big)$ is a decreasing function of $x_i$, $v_i(x_i)$ is also a decreasing function of $x_i$ according to \eqref{eq:derivative g wrt x rd}.
As in Section \ref{sec:fixed}, the increment rate $v_i(x_i)$ for all contents $i\in \{1,\cdots,F\}$ can be shown as Fig. \ref{fig:increase rate}. 
As a predetermined threshold $\mu$ decreases, $x_i=v_i^{-1}(\mu)$ gradually increases, and we can find the optimal $\mu^*$ which satisfies the constraint \eqref{eq:cache_const}, i.e., $\sum_{i=1}^F x^*_i=M/A$, where $x_i^* = v_i^{-1}(\mu^*)$ is optimal.
However, unlike Section \ref{sec:fixed}, $v_i^{-1}(x_i)$ cannot be derived analytically.
Thus, we use a bisection method to find $x^*_i$ satisfying $v_i(x^*_i)=\mu^*$.
The method of finding $x^*_i$ is presented in Algorithm \ref{alg:water-filling 2}.
Finally, we can obtain the optimal caching fraction $\alpha^*_i$ as $\alpha^*_i=x^*_i/q_{\max}$ using Lemma \ref{lemma:lemma2}.

\begin{algorithm}[t]
	\caption{Algorithm to find $\left\{x^*_i\right\}$ for the uniformly distributed device}\label{alg:water-filling 2}
	\begin{algorithmic}
		\State \textbf{Initialize} $a=0$ and $b=v_1(0)$
		\While{$b-a \geq \epsilon$}
		\State{$\mu=\frac{a+b}{2}$}
		\For{$i=1,\cdots,F$}
		\If{$\mu>v_i(0)$}
		\State $x_i=0$
		\ElsIf{$\mu \leq v_i(0)$}
		\State Find $x_i$ satisfying that $v_i(x_i)=\mu$ using \State the Bisection method
		\EndIf
		\EndFor
		\If{$\sum_{i=1}^F x_i > M/A$}
		\State $a=\mu$
		\ElsIf{$\sum_{i=1}^F x_i \leq M/A$}
		\State $b=\mu$
		\EndIf
		\EndWhile
	\end{algorithmic}
\end{algorithm}
\vspace{-2mm}

\vspace{-2mm}
\section{Numerical Results}
This section provides several numerical results to illustrate the impact of the proposed caching policies. %by comparing it with the traditional caching policy that stores the entire contents, not partial ones.
No matter what the distribution of devices is, the average distance from the BS is $\mathbb{E}[r]=40$ m and the transmission SNR is $\Psi=25\text{dB}$, unless otherwise noted.
The number of contents is $F=20$ and popularity $\left\{f_i\right\}$ is assumed to follow the Zipf distribution with the exponent of 1.
Also, $q_{\min}=0.2$ Mbps, $q_{\max}=1$ Mbps, $M=650 \text{kB}$, $\mathcal{B}=5\text{MHz}$, $\Upsilon=5$ dB, $\beta=3.0$, $T=1$ sec, and $A=1$ sec are used. 
%We set the path loss exponent as $\beta=3.0$ and the INR as $\Upsilon=5$ dB.
%The bandwidth is $\mathcal{B}=5\text{MHz}$.
%The time period and file size coefficient are assumed to be 
%Also, $T=1\text{sec}$ and $A=1$ sec are used. 
%The storage size of the device is assumed to be $M=650 \text{kB}$.
%The minimum and maximum quality are set to $q_{\min}=0.2$ Mbps and $q_{\max}=1$ Mbps, respectively.

\begin{figure}[t]
	\centering
	\includegraphics[width=0.35\textwidth, trim=-0.5cm 0 0 0]{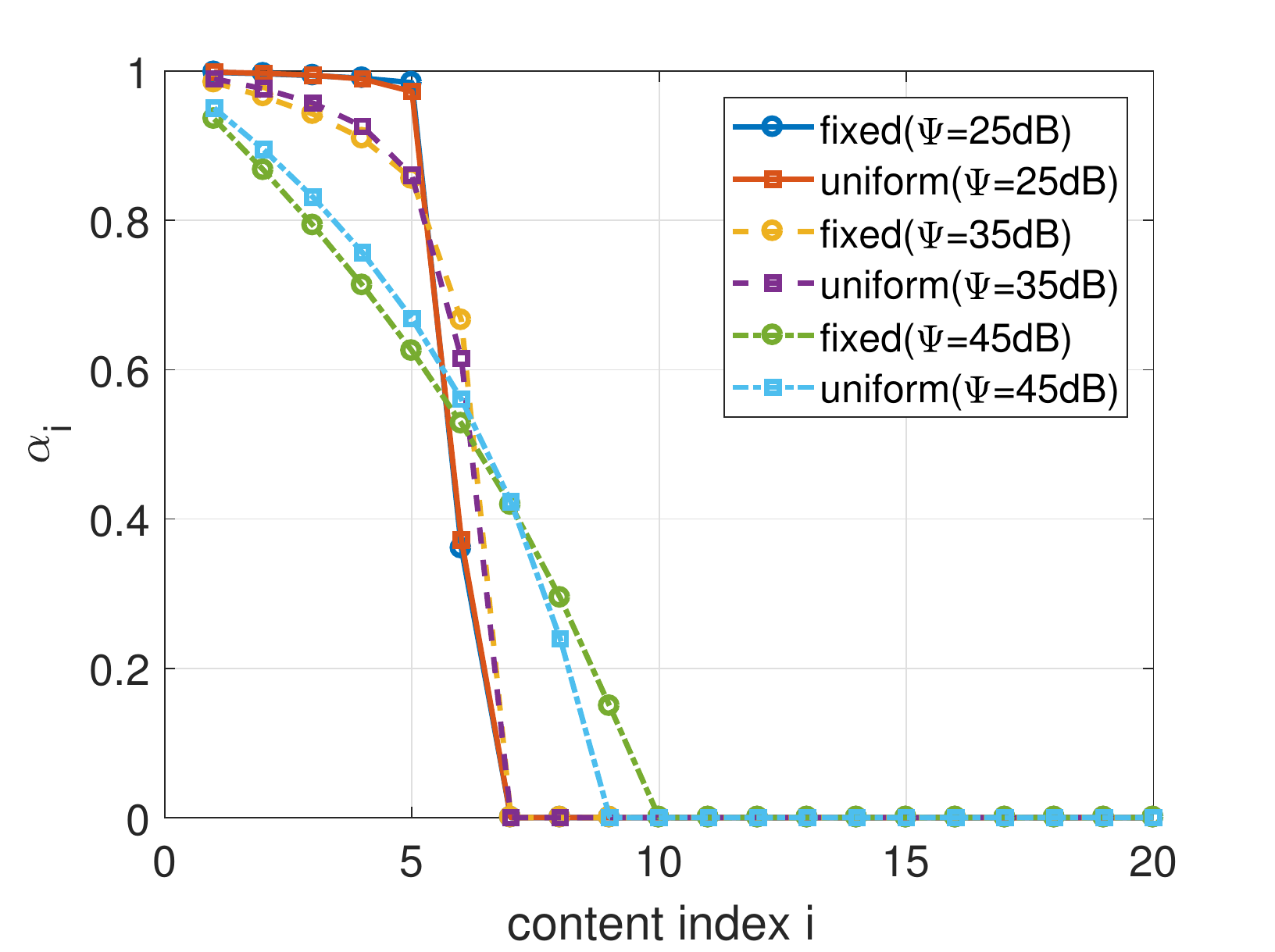}
	\caption{$\alpha_i$ versus content $i$ for various transmit SNR.}\label{fig:alpha_snr}
\end{figure}

\begin{figure}[t]
	\centering
	\includegraphics[width=0.35\textwidth, trim=-0.5cm 0 0 0]{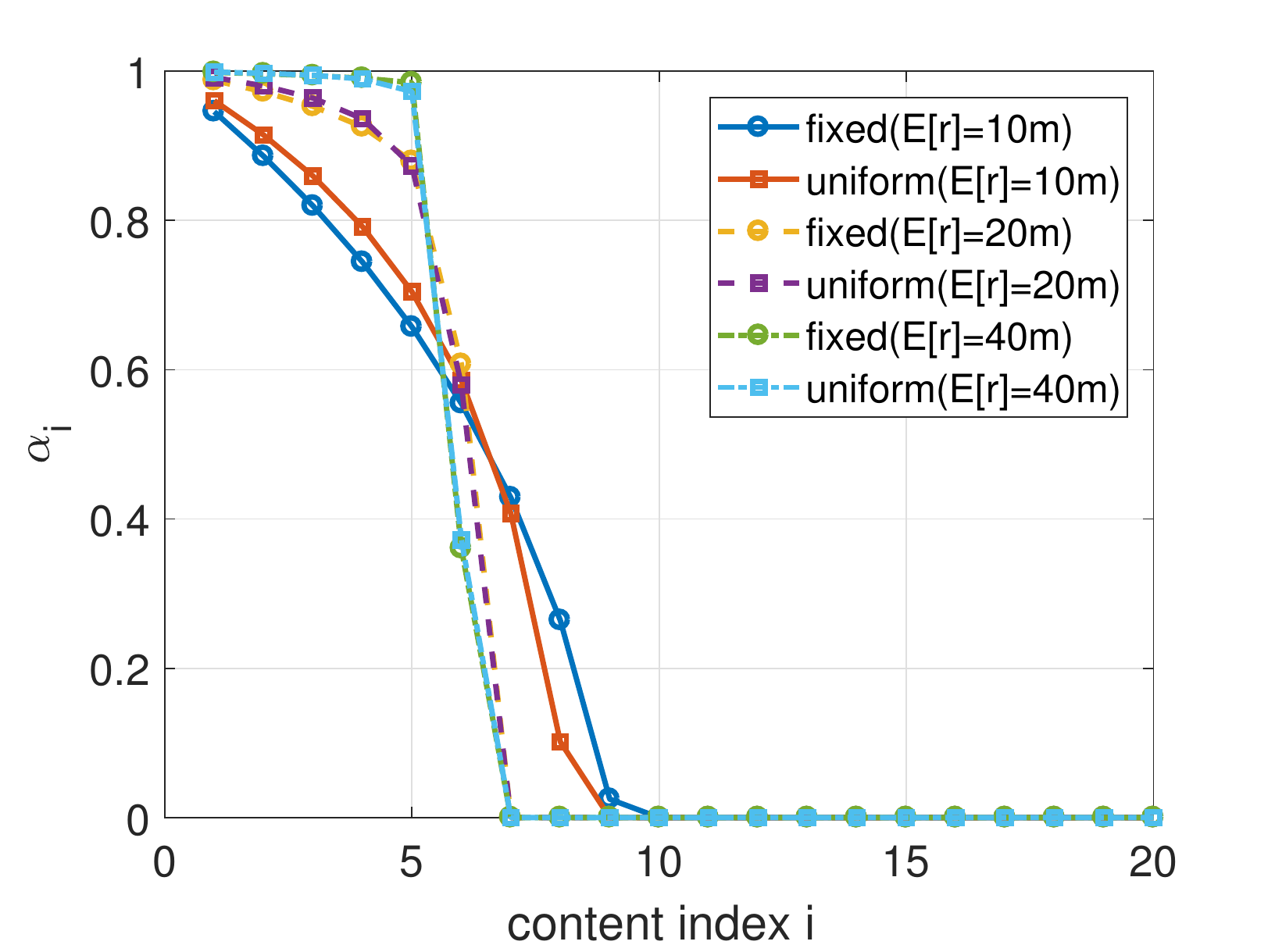}
	\caption{$\alpha_i$ versus content $i$ for various average distances.}\label{fig:alpha_d}
\end{figure}

Figs. \ref{fig:alpha_snr} and \ref{fig:alpha_d} show the plots of $\alpha_i$ versus the content index $i$ for various transmit SNRs and average distances, respectively. 
As the transmit SNR decreases, a content-requesting device becomes difficult to receive the uncached content fractions so that caching is biased; in other words, contents with high popularity are almost entirely cached and dominate the cache size. 
Meanwhile, when the SNR is large, the BS can successfully transmit the content fractions; therefore, the device caches relatively many contents by partitioning contents. 
However, no matter how good the channel is, an outage event can occur, so the more popular the content is, the more fractions are cached.
%It can be seen that the larger the transmit SNR, the more caching fraction for popular content.
%If the transmit SNR is large, the channel condition is good, so that the BS can transmit with the maximum quality, so only popular contents are cached.
%When the transmit SNR is low, BS cannot transmit enough. Therefore, caching is performed in proportion to popularity.
%In Fig. \ref{fig:alpha_d}, the plots of $\alpha_i$ versus content index $i$ for various average distances are described.
In Fig. \ref{fig:alpha_d}, a small average distance means that a content-requesting device can easily receive the uncached fractions from the BS; therefore, its effects on the caching policy are similar to those of a large SNR, which caches fractions of many contents. 
%It can be seen that caching is performed for contents with high popularity because the channel environment is improved as $\mathbb{E}[r]$ is small.

\begin{figure}[t]
	\centering
	\includegraphics[width=0.35\textwidth, trim=-0.5cm 0 0 0]{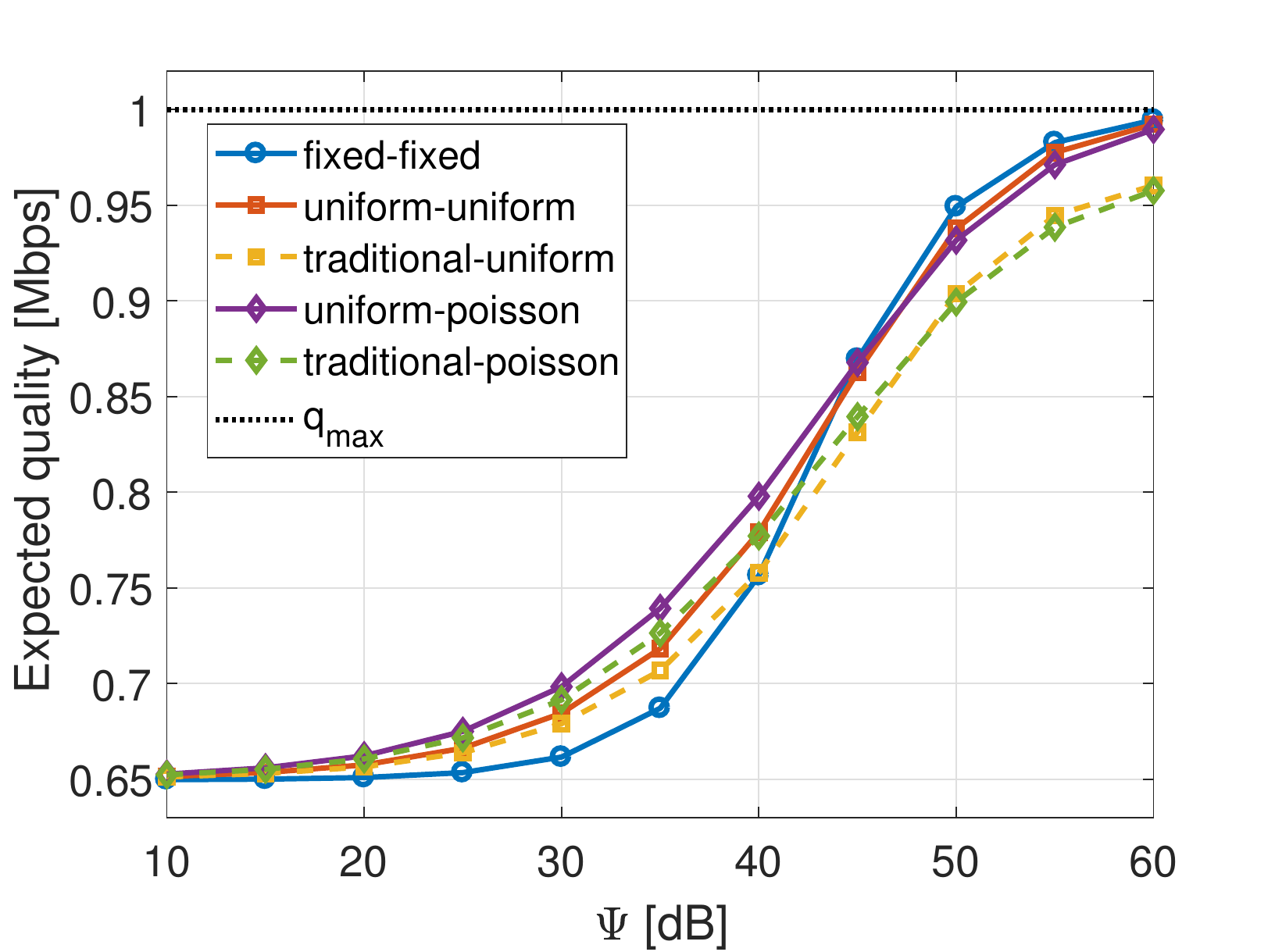}
	\caption{Expected quality versus transmit SNR.}\label{fig:exp_qual_snr}
\end{figure}

\begin{figure}[t]
	\centering
	\includegraphics[width=0.35\textwidth, trim=-0.5cm 0 0 0]{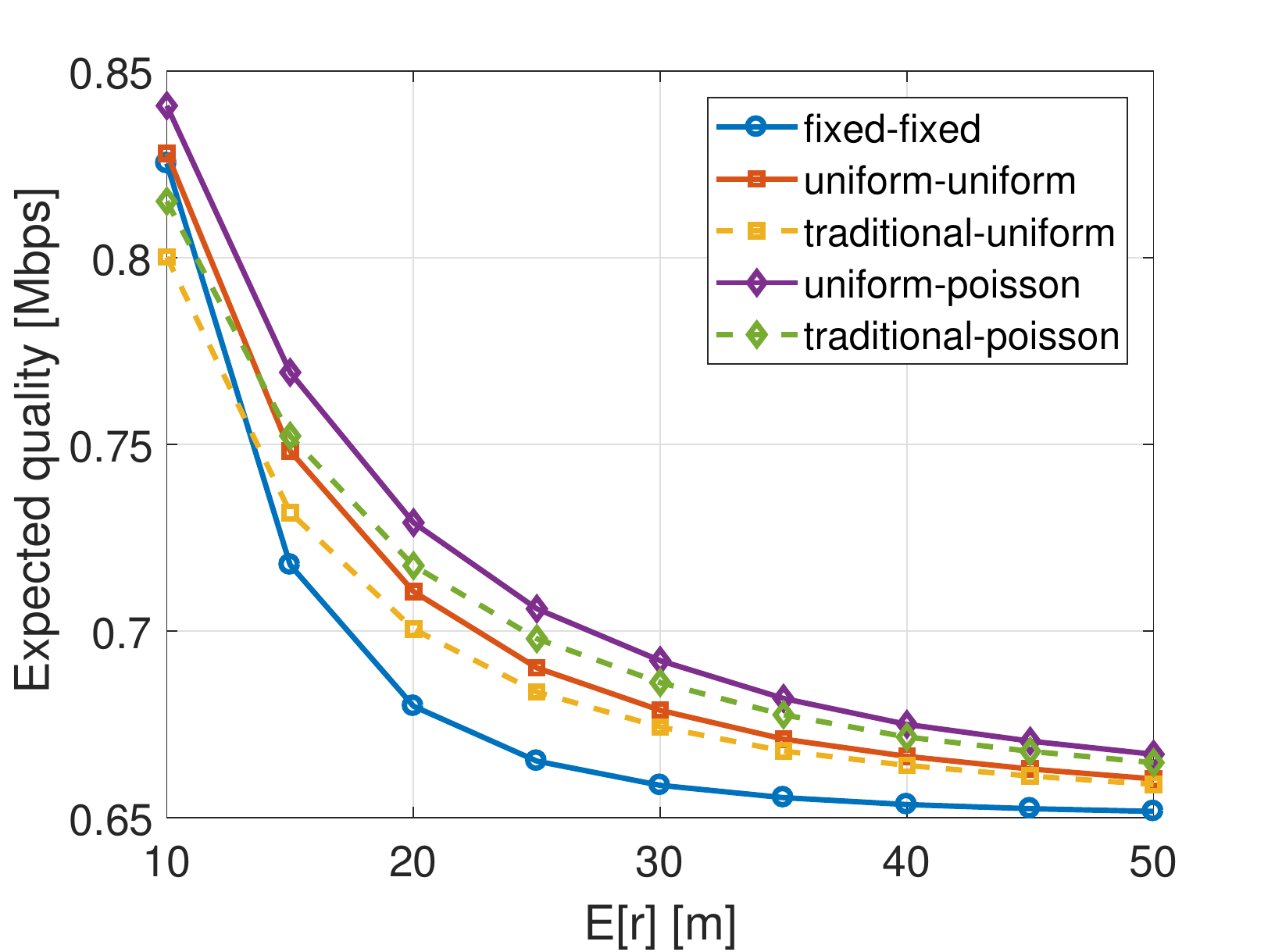}
	\caption{Expected quality versus average distance .}\label{fig:exp_qual_d}
\end{figure}

In Figs. \ref{fig:exp_qual_snr} and \ref{fig:exp_qual_d}, the proposed caching policy is compared with the traditional one that allows caching of entire contents only, in terms of the expected quality.
We perform simulations in three different device distributions: 1) fixed, 2) Uniform, and 3) Poisson, where 'fixed' indicates that a device is static and its distance from the BS is fixed, and 'Uniform' and 'Poisson' mean that the device location follows uniform and Poisson distributions, respectively. 
A legend '\textit{A}-\textit{B}' means that the caching policy optimized for \textit{A} distribution is applied to the environment in which devices are distributed under \textit{B} distribution.
In Fig. \ref{fig:exp_qual_snr}, parameters of each distribution are set to have the same average distance as that of 'fixed'.
In all cases, the higher the transmit SNR, the better the expected quality performance.
The proposed caching policy of content fraction shows better performance than the traditional caching policy in all distributions, and the advantage grows as the SNR increases.  
%Fig. \ref{fig:exp_qual_snr} shows the the expected quality versus transmit SNR for various caching policies and distributions.
%The traditional non-partitioned caching policy implies full caching of only popular contents as storage size permits.
%We perform simulations in three distance distribution environments 'Fixed', 'Uniform' and 'Poisson'.
%The parameters are set to have the same average distance in each distribution.
%Legend 'A-B' means that the caching policy optimized for A distribution is applied to B distributed environment.
%In all cases, the higher the transmit SNR, the better the expected quality performance.
%They have an upper bound $q_{\max}$, and they get closer as the transmit SNR gets higher.
%The proposed caching policy of content fraction shows better performance than the traditional caching policy in all distributions.
Fig. \ref{fig:exp_qual_d} shows the expected quality as a function of average distance.
A large distance between the device and the BS makes difficult to deliver the uncached fractions from the BS to the device; therefore, the advantage of the proposed caching policy decreases. 
%As the distance increases, the channel condition deteriorates, so the performance also decreases.
%As in Fig. \ref{fig:exp_qual_snr}, it can be seen that the proposed caching policy outperforms the traditional caching policy.

%Thus, we can conclude that when the channel capacity of the link between a content-requesting user and any node that can provide the requested content to the user is sufficiently good, partitioning the content and caching of content fractions would be better than caching of contents entirely. 
%A D2D-assisted caching network in which a user can easily find several cache-enabled devices nearby motivates caching of content fractions, and this work would be the first guideline.

%\vspace{-3mm}

%\vspace{-3mm}
\section{Conclusion}
In this paper, we consider caching of contents that can be partitioned into many fractions and have different qualities, and analyze the impact of caching content fractions in a user device on the expected quality.
%the trade-off between the quality of content and the caching fraction in a network where device caching is possible, and propose the optimal caching policy that maximizes the expected quality of user.
Numerical results show that the proposed caching policy outperforms the conventional policy that caches contents entirely, especially when channel capacity of the link between a content-requesting user and any node that can deliver the requested content to the user is sufficiently good. 
A D2D-assisted caching network in which a user can easily find several cache-enabled devices nearby motivates caching of content fractions, and this work would be the first guideline.

%\vspace{-4mm}
%\section*{Acknowledgment}
%This work was supported by the National Research Foundation of Korea under Grant NRF-2020R1G1A1101164.

\bibliographystyle{IEEEtran.bst}
\bibliography{wcl2021.bib}

\begin{thebibliography}{10}
\providecommand{\url}[1]{#1}
\csname url@samestyle\endcsname
\providecommand{\newblock}{\relax}
\providecommand{\bibinfo}[2]{#2}
\providecommand{\BIBentrySTDinterwordspacing}{\spaceskip=0pt\relax}
\providecommand{\BIBentryALTinterwordstretchfactor}{4}
\providecommand{\BIBentryALTinterwordspacing}{\spaceskip=\fontdimen2\font plus
\BIBentryALTinterwordstretchfactor\fontdimen3\font minus
  \fontdimen4\font\relax}
\providecommand{\BIBforeignlanguage}[2]{{%
\expandafter\ifx\csname l@#1\endcsname\relax
\typeout{** WARNING: IEEEtran.bst: No hyphenation pattern has been}%
\typeout{** loaded for the language `#1'. Using the pattern for}%
\typeout{** the default language instead.}%
\else
\language=\csname l@#1\endcsname
\fi
#2}}
\providecommand{\BIBdecl}{\relax}
\BIBdecl

\bibitem{TITShanmugam}
K.~Shanmugam, N.~Golrezaei, A.~G. Dimakis, A.~F. Molisch, and G.~Caire,
  ``Femtocaching: Wireless content delivery through distributed caching
  helpers,'' \emph{IEEE Transactions on Information Theory}, vol.~59, no.~12,
  pp. 8402--8413, 2013.

\bibitem{TMM2013Cheng}
X.~Cheng, J.~Liu, and C.~Dale, ``Understanding the characteristics of internet
  short video sharing: A youtube-based measurement study,'' \emph{IEEE
  Transactions on Multimedia}, vol.~15, no.~5, pp. 1184--1194, 2013.

\bibitem{JSAC2016Ji}
M.~Ji, G.~Caire, and A.~F. Molisch, ``Wireless device-to-device caching
  networks: Basic principles and system performance,'' \emph{IEEE Journal on
  Selected Areas in Communications}, vol.~34, no.~1, pp. 176--189, 2016.

\bibitem{JSAC2018Choi}
M.~Choi, J.~Kim, and J.~Moon, ``Wireless video caching and dynamic streaming
  under differentiated quality requirements,'' \emph{IEEE Journal on Selected
  Areas in Communications}, vol.~36, no.~6, pp. 1245--1257, 2018.

\bibitem{TMC2019Chat}
L.~E. Chatzieleftheriou, M.~Karaliopoulos, and I.~Koutsopoulos, ``Jointly
  optimizing content caching and recommendations in small cell networks,''
  \emph{IEEE Transactions on Mobile Computing}, vol.~18, no.~1, pp. 125--138,
  2019.

\bibitem{TWC2021Choi}
M.~Choi, A.~F. Molisch, D.-J. Han, D.~Kim, J.~Kim, and J.~Moon, ``Probabilistic
  caching and dynamic delivery policies for categorized contents and
  consecutive user demands,'' \emph{IEEE Transactions on Wireless
  Communications}, vol.~20, no.~4, pp. 2685--2699, 2021.

\bibitem{TMC2020Qu}
Z.~Qu, B.~Ye, B.~Tang, S.~Guo, S.~Lu, and W.~Zhuang, ``Cooperative caching for
  multiple bitrate videos in small cell edges,'' \emph{IEEE Transactions on
  Mobile Computing}, vol.~19, no.~2, pp. 288--299, 2020.

\bibitem{TMC2019Tran}
T.~X. Tran and D.~Pompili, ``Adaptive bitrate video caching and processing in
  mobile-edge computing networks,'' \emph{IEEE Transactions on Mobile
  Computing}, vol.~18, no.~9, pp. 1965--1978, 2019.

\bibitem{TWC2019Choi}
M.~Choi, A.~No, M.~Ji, and J.~Kim, ``Markov decision policies for dynamic video
  delivery in wireless caching networks,'' \emph{IEEE Transactions on Wireless
  Communications}, vol.~18, no.~12, pp. 5705--5718, 2019.

\bibitem{JCN2021Choi}
M.~Choi, M.~Shin, and J.~Kim, ``Dynamic video delivery using deep reinforcement
  learning for device-to-device underlaid cache-enabled internet-of-vehicle
  networks,'' \emph{Journal of Communications and Networks}, vol.~23, no.~2,
  pp. 117--128, 2021.

\bibitem{NCC2019Narayana}
V.~C.~L. Narayana, S.~Jain, and S.~Moharir, ``Caching partial files for content
  delivery,'' in \emph{2019 National Conference on Communications (NCC)}, 2019,
  pp. 1--6.

\bibitem{TWC2020Luo}
J.~Luo, F.~R. Yu, Q.~Chen, and L.~Tang, ``Adaptive video streaming with edge
  caching and video transcoding over software-defined mobile networks: A deep
  reinforcement learning approach,'' \emph{IEEE Transactions on Wireless
  Communications}, vol.~19, no.~3, pp. 1577--1592, 2020.

\end{thebibliography}

\end{document}